\documentclass[amsmath,amssymb,prl,aps,twocolumn,mathbbm,superscriptaddress]{revtex4}
\usepackage{amssymb}
\usepackage{amsmath}
\usepackage{graphicx}
\usepackage{times}
\usepackage{subfigure}
\usepackage{amstext}
\usepackage{amsfonts}
\usepackage[colorlinks=true, citecolor=blue]{hyperref}

\usepackage[usenames]{color}

\begin{document}
\title{Quantum Communication Between Remote Mechanical Resonators}
\author{S. Felicetti}
\address{Laboratoire Mat\' eriaux et Ph\' enom\`enes Quantiques, Sorbonne Paris Cit\' e, Universit\' e Paris Diderot, CNRS UMR 7162, 75013, Paris, France}
\author{S. Fedortchenko}
\address{Laboratoire Mat\' eriaux et Ph\' enom\`enes Quantiques, Sorbonne Paris Cit\' e, Universit\' e Paris Diderot, CNRS UMR 7162, 75013, Paris, France}
\author{R. Rossi Jr.}
\address{Universidade Federal de Vi\c{c}osa - Campus Florestal, LMG818 Km6, Minas Gerais, Florestal 35690-000, Brazil}
\author{S. Ducci}
\address{Laboratoire Mat\' eriaux et Ph\' enom\`enes Quantiques, Sorbonne Paris Cit\' e, Universit\' e Paris Diderot, CNRS UMR 7162, 75013, Paris, France}
\author{I. Favero}
\address{Laboratoire Mat\' eriaux et Ph\' enom\`enes Quantiques, Sorbonne Paris Cit\' e, Universit\' e Paris Diderot, CNRS UMR 7162, 75013, Paris, France}
\author{T. Coudreau}
\address{Laboratoire Mat\' eriaux et Ph\' enom\`enes Quantiques, Sorbonne Paris Cit\' e, Universit\' e Paris Diderot, CNRS UMR 7162, 75013, Paris, France}
\author{P. Milman}
\address{Laboratoire Mat\' eriaux et Ph\' enom\`enes Quantiques, Sorbonne Paris Cit\' e, Universit\' e Paris Diderot, CNRS UMR 7162, 75013, Paris, France}
\date{\today}

\begin{abstract}
Mechanical resonators represent one of the most promising candidates to mediate the interaction between different quantum technologies, bridging the gap between efficient quantum computation and long-distance quantum communication. In this letter, we introduce a novel interferometric scheme where the interaction of a mechanical resonator with input/output quantum pulses is controlled by an independent classical drive. We design protocols for state teleportation and direct quantum state transfer, between distant mechanical resonators. The proposed device, feasible with state-of-the-art technology, can serve as building block for the implementation of long-distance quantum networks of mechanical resonators.

\end{abstract}

\pacs{}
\maketitle

During the spectacular development of quantum technologies observed in the last decades, it has become clear that no single platform will emerge as the best candidate to perform the totality of all
quantum information tasks~\cite{Ladd10}. For example, superconducting circuits~\cite{Devoret13} and atomic systems~\cite{Bloch08,Blatt12} excel in quantum computation and quantum simulation.  Photonic systems~\cite{Brien09}, among many other achievements, have proven able to implement quantum communication over unchallenged distances. The feasibility of building quantum systems of increasing size and complexity will strongly depend on our ability to develop a hybrid approach~\cite{Pirandola16}, exploiting different quantum technologies in large-scale quantum networks~\cite{Kimble08}. 
In this framework, micro- and nano-mechanical devices come through as potential links between different platforms, as they can be coupled to a plethora of different quantum systems~\cite{Wallquist09,Treutlein14,Aspelmeyer14}.
So far, it has been experimentally explored the interaction of mechanical oscillators with ensembles of cold atoms~\cite{Treutlein07}, microwave and optical cavities~\cite{Favero09, Metzger08, Ding10, Nunnenkamp11, Verhagen12,Palomaki13, Krause15, Riedinger16}, superconducting qubits~\cite{Connell10}, and single spins~\cite{Kolkowitz12}. Using optomechanical devices, coherent conversion of microwave to optical photons~\cite{Hill12, Bochmann13}, and viceversa~\cite{Andrews14}, has already been demonstrated.

The interest in using mechanical resonators in quantum communication protocols has been steadily growing. Most recently, the interaction of mechanical resonators with travelling optical signals has been studied for the purpose of light-to-mechanical teleportation~\cite{Hofer11}, qubit-to-light transduction~\cite{Stannigel10}, non-Gaussian state swapping~\cite{Filip15}, and entanglement generation~\cite{Hofer15, Vivoli16}. Different schemes have been proposed to distribute entanglement among remote mechanical resonators~\cite{Abdi12, Vostrosablin16}. However, a complete protocol for transferring arbitrary quantum states between distant mechanical resonators is still missing. 
In general, interconnecting remote quantum systems through propagating quantum signals is a highly challenging task. Quantum state transfer and teleportation have been realized for distant single atoms in a probabilistic way \cite{Nolleke13,Ritter12}. Quantum state teleportation has been implemented in a probabilistic fashion between distant trapped ions \cite{Olmschenk09}, as well as in a deterministic way between spatially separated atomic clouds \cite{Krauter13}.

In this letter, we propose a novel interferometric scheme, where a classical pump amplifies and controls the interaction between a mechanical resonator and propagating quantum optical signals. By tuning in real-time the pump frequency and intensity, it is possible to change the nature of the optomechanical interaction, and to control the temporal shape of input/output optical pulses.  
This feature enables us to overcome  the issue of temporal-envelope mismatch in the absorption and emission processes.
We design protocols  to implement mechanical-to-mechanical continuous-variable  teleportation, or to directly transmit arbitrary states between two remote mechanical resonators over a quantum optical channel. The proposed scheme is implementable with state-of-the-art technology and it works at telecommunication wavelengths.

\begin{figure}[t]
\centering
\includegraphics[angle=0, width=0.35\textwidth]{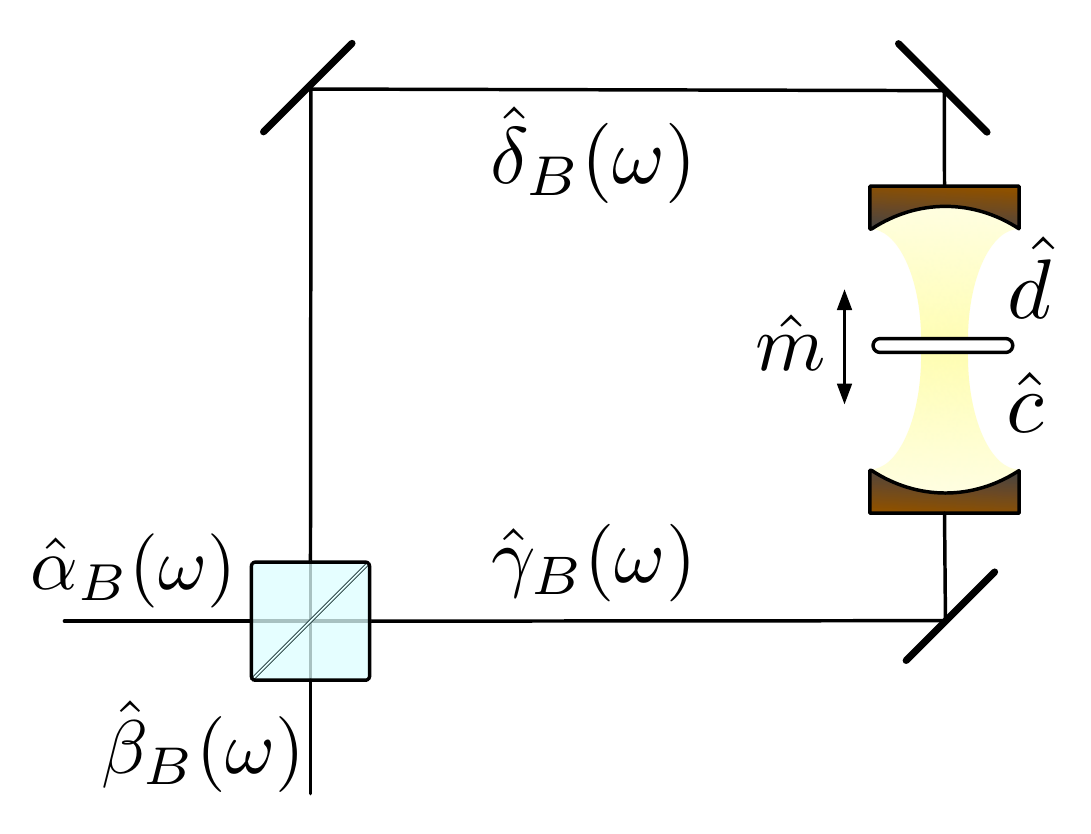}
\caption{\label{sketch} Sketch of the proposed interferometric scheme. An optomechanical device, composed of two identical optical cavities sharing a vibrating mirror, is embedded in a common-path Sagnac interferometer. In our protocol, a classical drive sent through the control port $\hat \beta_B(\omega)$ modulates the optomechanical interactions with input/output quantum signals $\hat \alpha_B(\omega)$.}   
\end{figure}

We consider a system composed of two resonant single-mode optical cavities, which share a mirror made of a non-transmissive vibrating membrane. We assume that the membrane effectively supports a single phononic mode, since intermode couplings are typically negligible. The cavity is embedded in a symmetric Sagnac interferometer, as shown in Fig.~\ref{sketch}. The radiation pressure induces an optomechanical coupling between the optical modes $\hat c$, $\hat d$ and the vibrational mode $\hat m$, such that the system Hamiltonian can be written as
\begin{equation}
\hat H =  \hat H_0   - g_0 \left( \hat {d}^\dagger \hat d - \hat c^\dagger \hat c \right) \left( \hat m^\dagger + \hat m \right),
\end{equation}
where $\hat H_0 = \omega_c \left( \hat d^\dagger\hat d + \hat c^\dagger \hat c\right) + \omega_m \hat m^\dagger \hat m$. We denote with $\omega_c$ and $\omega_m$ the frequencies of the optical cavities and mechanical modes, respectively, and with $g_0$ the optomechanical coupling strength, assumed to be equal for the two modes. To derive input/output relations for our system, we consider a standard Markovian coupling of the intra-cavity fields $\hat c$ and $\hat d$ with the modes describing the 
electromagnetic environment, $\hat \gamma_{B}(\omega)$ and $\hat \delta_{B}(\omega)$, respectively. It is straightforward to see that the input modes of the 50/50 beam splitter, $\hat \alpha_{B}(\omega)$ and $\hat \beta_{B}(\omega)$ in Fig.\ref{sketch},  interact selectively with the bosonic modes $\hat a$ and $\hat b$, defined by the relations
$\hat a = \left(\hat c -\hat d \right)/\sqrt{2}$  and  $b = \left(\hat c +\hat d \right)/\sqrt{2}$. These modes correspond to collective excitations of the two optical intra-cavity modes. 
The system Hamiltonian can thus be rewritten as
\begin{equation}
\hat H =  \hat H_0   + g_0 \left(\hat  a^\dagger \hat b +\hat a \hat b^\dagger \right)\left(\hat m^\dagger +\hat m \right), \label{H2}
\end{equation}
where $\hat H_0 = \omega_c \left(\hat a^\dagger \hat a + \hat b^\dagger \hat b\right) + \omega_m \hat m^\dagger \hat m$. Let us now assume that an undepleted coherent state at frequency $\omega_b$ is sent through the port $\hat \beta_{B}(\omega)$. The spatial mode $\hat \beta_{B}(\omega)$ can be decomposed in two modes propagating in opposite directions, proportional to the Fourier transforms of the $\hat b$ mode's input and output respectively \cite{Gardiner,GardinerBook,WallsBook}. We can describe the driven mode $b$  as a classical field, and replace $\hat b\rightarrow \beta \, e^{-i \omega_{b} t}$ in Eq.~(\ref{H2}). In a frame where $a$ is rotating at the drive frequency, the Hamiltonian becomes (see Supplemental Material)
 \begin{equation}
\hat H_\text{eff} = \hat H_{0}' + g_0\beta
 \left(\hat a^\dagger +  \hat a \right)\left(\hat m^\dagger +\hat m \right),
\label{Hlin}
\end{equation}
where $\hat H_0' = \left(\omega_c - \omega_{b} \right)\hat a^\dagger \hat a + \omega_m \hat m^\dagger \hat m$, where we assumed $\beta$ real.
Notice that this derivation leads to a linear optomechanical interaction of tunable strength $g_0 \beta$. In the context of linearized optomechanics~\cite{Aspelmeyer14}, such coupling is usually obtained considering small quantum fluctuations on top of a classical signal, which enhances the otherwise negligible interaction~\cite{Liao15}.
In our scheme, the quantum input and the classical pump are given by independent modes, a fundamental difference that presents various advantages.
1) The Hamiltonian in Eq.~(\ref{Hlin}) is valid for large quantum fluctuations of the intra-cavity field, where the standard linear approximation of optomechanics breaks down. 2) The proposed optomechanical device is able to interact with arbitrary quantum inputs, without requiring displacing operations beforehand or additional non-linear elements. 3) The effective coupling strength $g_0 \beta$ and the coupling frequency $\omega_b$ depend on a classical drive, which is independent on the state and frequency of the quantum input. 4) The quantum and the classical output exit the device via independent optical paths.

The proposed scheme can be used as a black-box with an internal quantum variable (the mechanical resonator), a quantum input/output port $\alpha_{B}$ and a classical control port $\beta_{B}$ that modulates the interaction between them. We consider the resolved-sideband regime $\kappa\ll \omega_m$, where $\kappa$ is the cavity dissipation rate. Under the requirement that the effective coupling strength is smaller than the mode frequencies $g_{0}\beta \ll \omega_m, \omega_c$, tuning $\omega_b$ enables the selective activation of the red or the blue sideband of the optomechanical interaction. We consider a long-pulsed regime~\cite{Hofer11}, such that the total interaction time $\tau$ is short compared with the timescale $\gamma$ of mechanical dissipative processes, but long enough to adiabatically eliminate the cavity mode $\kappa^{-1}\ll \tau \ll \gamma^{-1}$.

Let us consider first the case in which  the red sideband is selected  in Eq.~\eqref{Hlin}, $i.e.$, the interaction terms $\hat a^\dagger \hat m +  \hat a \, \hat m^\dagger$. The classical drive is a coherent pulse of frequency
$\omega_b = \omega_a - \omega_m$, of duration $\tau$. In the interaction picture, the Langevin equations are given by \cite{Gardiner,GardinerBook,WallsBook},
\begin{eqnarray}
\label{TRANSlan}
\dot {\hat a}(t) &=& - \kappa   \hat a(t) -i g\hat m (t) - \sqrt{2 \kappa }\ \hat a_{in}(t), \\
\dot {\hat m}(t) &=& - \gamma   \hat m(t) -ig \hat a(t) - \sqrt{2 \gamma }\ \hat m_{in} (t) . \nonumber
\end{eqnarray}
Performing an adiabatic elimination of the cavity mode (see Supplemental Material), we are able to find analytically the input/output relations after a fixed interaction time $\tau$,
\begin{eqnarray}
\label{redinout}
\hat A^r_{out} &=& -  e^{-G\tau}\hat A^r_{in} - i\sqrt{ 1-  e^{-2G\tau} }\ \hat M_{in}, \\
\hat M_{out} &=& \ e^{-G\tau} \hat M_{in}  + i\sqrt{ 1-  e^{-2G\tau}  }\  \hat A^r_{in} - C^r \hat M^r_B, \nonumber
\end{eqnarray}
where $G = \frac{(g_0\beta)^2}{\kappa }$.
We defined the mechanical state before $\hat M_{in} = m(0)$  and after $\hat M_{out} =\hat m(\tau)$ the interaction. The relations of Eq.~\eqref{redinout} then  describe a state-swap process between the mechanical resonator and the normalized temporal modes $\hat A^r_{in} =   \mathit{ Q}(\tau,\hat a_{in} )$ and $\hat A^r_{out} = \mathit{P}(\tau,\hat a_{out} )$, which are defined by the functions,
\begin{eqnarray}
\label{PQdef}
\mathit{P}(\tau,\hat O) = \sqrt{\frac{2G}{1-e^{-2G\tau}}}\int_0^\tau dt e^{-Gt}\hat O (t),\\
 \mathit{ Q}(\tau,\hat O ) = \sqrt{\frac{2G}{e^{2G\tau}-1}}\int_0^\tau dt e^{Gt}\hat O (t). \nonumber
\end{eqnarray}
For the state-swap interaction to take place, the input/output pulses must be modulated by a specific exponential envelope given by Eq.~\eqref{PQdef}.
The interaction of the mechanical resonator with the thermal environment is included in the model via the term $C^r \hat M^r_B$, and it will set the upper bound to the optimal interaction time for the state-swap process. We defined $C^r = \sqrt{\frac{\gamma }{G}\left(1- e^{-2G\tau}\right)}$, and 
$\hat M^r_B = \mathit{ Q}(\tau,\hat m_{in} )$, being $\hat m_{in}$ the standard input operator for bosonic bath modes. The impact of the mechanical bath on the optical output is a second-order effect in $\gamma /G$, and it is negligible for the interaction times $\tau$ here considered.

\begin{figure}[h]
\centering
\includegraphics[angle=0, width=0.45\textwidth]{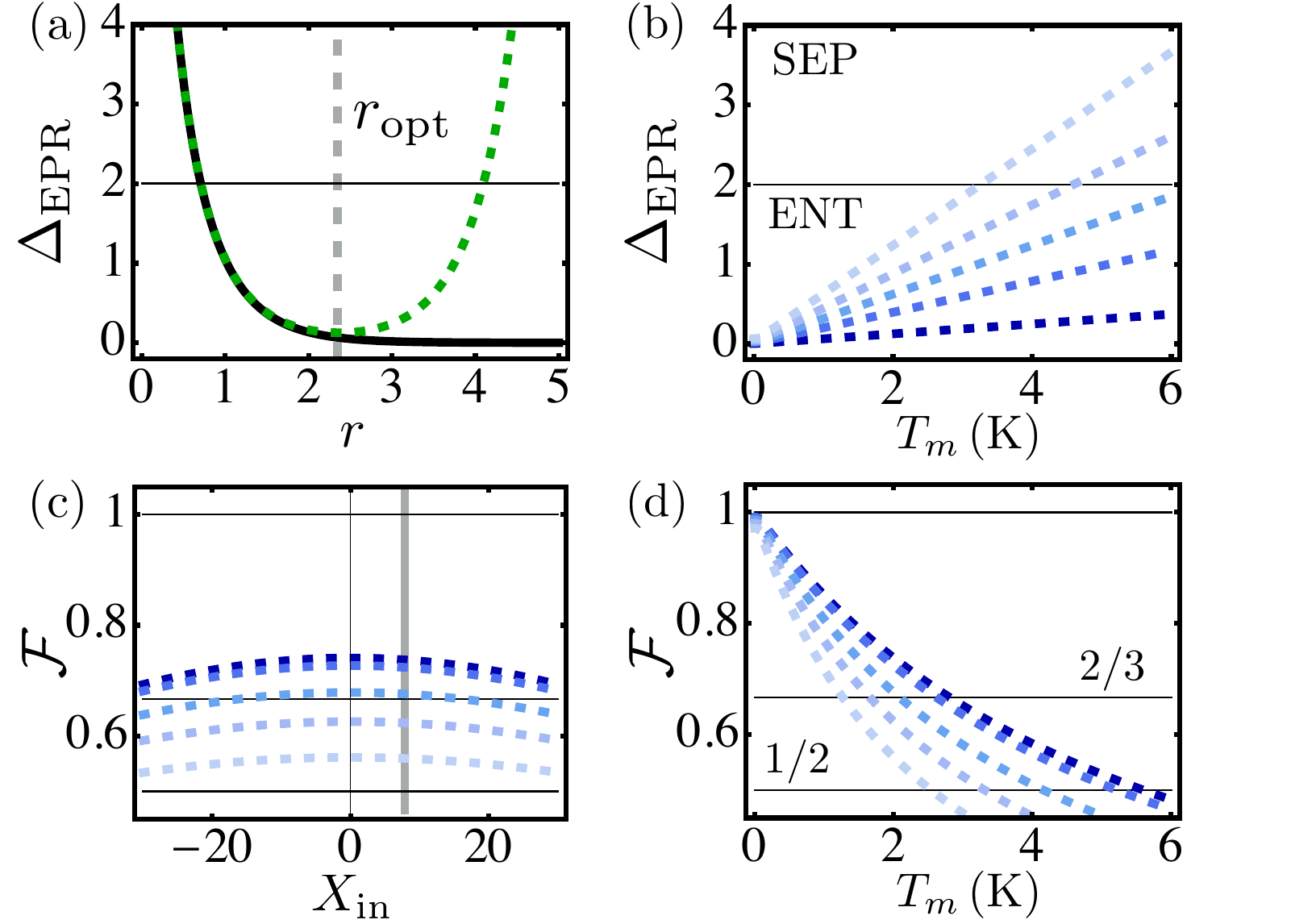}
\caption{\label{teleportF} Teleportation protocol. (a) EPR variance as a function of the squeezing parameter without (black full line) and with (green dashed line) mechanical dissipation,
as functions of the squeezing parameter $r=G \, \tau$. The state is entangled if $\Delta_{\text{EPR}} < 2$. Here $T_{m}=2$ K, $g=0.05 \, k_{c}$, $k_{c} = 0.1 \, \omega_{m}$, $\omega_{m} / k_{m}=10^{7}$, $\omega_{c} / k_{c}=10^{7}$, while $\omega_{m}/ 2\, \pi \sim 10^9$ Hz and $\omega_{c}/ 2\, \pi \sim 10^{15}$ Hz. The optimal squeezing parameter is given by $r_{\text{opt}} \approx 2.37$. (b) EPR variance as a function of the mechanical bath temperature. (c) Teleportation fidelity of a coherent state, as a function of its size $X_{\text{in}}$ (here $P_{\text{in}}=0$) for a displacement efficiency $\eta = 0.99$, with same parameters as in (a). The vertical gray line shows the value of $X_{\text{in}}$ used in (d). (d) Teleportation fidelity of a coherent state of size $X_{\text{in}} = \sqrt{50}$ ($P_{\text{in}}=0$), as a function of the mechanical bath temperature, for a displacement efficiency $\eta = 0.99$, with same parameters as in (a). (b)-(d) Color code: from darker to lighter shades of blue, $k_{m}$ ($r_{\text{opt}}$) is increased (decreased) as follows: $k_{m}=\omega_{m} / 10^{7}$ ; $\omega_{m} / 10^{6}$ ; $\omega_{m} 2.5 / 10^{6}$ ; $\omega_{m} 5 / 10^{6}$ ; $\omega_{m} / 10^{5}$, and $r_{\text{opt}} \approx 2.37$ ; $1.80$ ; $1.57$ ; $1.40$ ; $1.24$. }   
\end{figure}

On the other hand, when we set $\omega_b = \omega_a + \omega_m$,  the blue sideband in Eq.~\eqref{Hlin} is selected,  $i.e.$, the interaction terms $\hat a^\dagger \hat m^\dagger +  \hat a \, \hat m$. The corresponding Langevin equations are given by
\begin{eqnarray}
\label{Entlan}
\dot {\hat a}(t) &=& - \kappa   \hat a(t) -i g\hat m (t)^\dagger - \sqrt{2 \kappa }\ \hat a_{in}(t), \\
\dot {\hat m}(t) &=&  - \gamma \hat m(t) -ig \hat a(t)^\dagger  - \sqrt{2 \gamma }\ \hat m_{in} (t). \nonumber
\end{eqnarray}
Under the same assumptions, analytical solutions for the input/output relations can be found
\begin{eqnarray}
\label{blueinout}
\hat A^b_{out} &=& -  e^{G\tau}\hat A^b_{in} - i\sqrt{ e^{2G\tau}-1 }\ \left(\hat M_{in}\right)^\dagger, \\
\hat M_{out} &=& \ e^{G\tau} \hat M_{in}  + i\sqrt{ e^{2G\tau}-1 }\  \left(\hat A^b_{in}\right)^\dagger - C^b \hat M^b_B. \nonumber
\end{eqnarray}
This relation corresponds to an EPR state generation between the mechanical resonator and a travelling light mode defined by an exponentially shaped envelope, with central frequency $\omega_c$. In this case, the solutions are found in terms of the  normalized temporal modes $A^b_{in} =  \mathit{ P}(\tau,\hat a_{in} )$ and $A^b_{out} = \mathit{Q}(\tau,\hat a_{out} )$, defined in Eq.~\eqref{PQdef}. Notice that the time-envelope of the input/output pulses of the blue sideband process (entanglement generation) are the opposite ones with respect to the red sideband (state-swap). Here, the effect of the thermal environment is given by  the term $C^b \hat M^b_B$, where $C^b = \sqrt{\frac{\gamma }{G}\left(e^{2G\tau} -1\right)}$, and $\hat M^b_B = \mathit{ P}(\tau,\hat m_{in} )$. To assess the degree of entanglement of the  generated state we use the EPR variance~\cite{Duan00}, a figure of merit that can be smaller than 2 only for entangled states. The EPR variance is shown in Fig.~\ref{teleportF}(a)  as a function of the squeezing parameter $r=G \tau$, and in Fig.~\ref{teleportF}(b) as a function of temperature, for different mechanical dissipation rates.
 The optimal pulse duration is given by a trade-off between entanglement generation and mechanical dissipation rate.

\begin{figure}[]
\centering
\includegraphics[angle=0, width=0.35\textwidth]{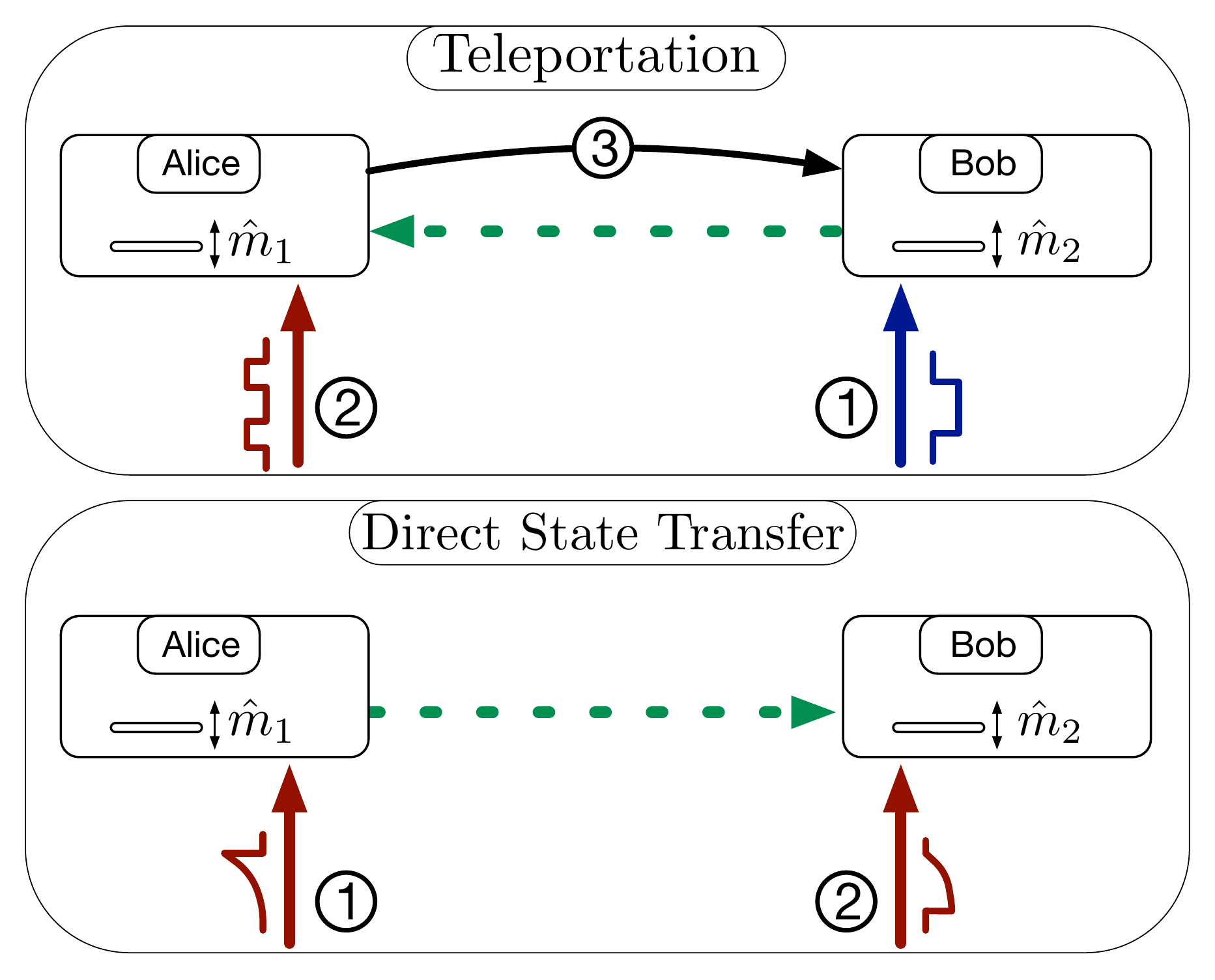}
\caption{\label{protocols} Schemes of the teleportation and state transfer protocols. Circled numbers represent different temporal steps. (Top panel) Teleportation protocol. (1) Bob generates an optomechanical EPR state. (2) Alice implements a Beam-Splitter interaction and the required measurements. (3) The results of the measurements is sent over a classical communication channel to Bob. (Bottom panel) State transfer process between remote mechanical oscillators. The classical drive must be time-modulated in order to optimize the time-envelope of the quantum signal.}   
\end{figure}

The red- and blue-sideband interactions are the fundamental operations we need in order to implement a mechanical-to-mechanical continuous-variable teleportation protocol~\cite{Braunstein98}. In the case in which the state to be teleported is encoded in Alice's device, the protocol is composed of the following steps, sketched in Fig~\ref{protocols}. 1) Through the blue sideband process, Bob generates an EPR state between its mechanical device and a light pulse, which is sent to Alice. 2) Alice implements a beam-splitter (BS) interaction between the received pulse and her mechanical oscillator. Such a BS interaction can be obtained with the red sideband process of Eq.~\eqref{redinout}, setting the effective coupling parameter $G$ so that $e^{-G \tau}=1/\sqrt{2}$. Notice that the time envelope of the EPR pulse generated in step 1 matches the optimal shape given by the input/output relations of Eq.~\eqref{redinout}. Then, the output optical pulse is measured via homodyne detection, while the mechanical resonator is measured through red-sideband interaction with a probe field. 3) The results of the two measurements are sent through a classical channel to Bob, who uses this information to choose the phase of a displacement operation to be applied to his mechanical resonator. At the end of the protocol, the state is destroyed in Alice's device and deterministically teleported onto Bob's.

Let us denote the quadratures of Bob's mechanical resonator with $\hat X_2(t) = \left[ \hat m_2(t)^\dagger +\hat m_2(t)\right]/\sqrt{2}$ and $\hat P_2(t) = i\left[ \hat m_2(t)^\dagger - \hat m_2(t)\right]/\sqrt{2}$ , and Alice's ones with $\hat X_1(t)$ and $\hat P_1(t)$. At the end of the teleportation protocol, the state of Bob's mechanical resonator is given by (see Supplemental Material)
\begin{eqnarray}
\label{teleportstate}
\hat X_2^{tel} = \eta \hat X_1(0) + R \hat X_2(0) + R^\prime \hat P_{in}, \\  \nonumber
\hat P_2^{tel} = \eta \hat P_1(0) - R \hat X_2(0) - R^\prime \hat X_{in},
\end{eqnarray}
where $\hat X_{in} =  \left[ (\hat A^b_{in}) ^\dagger +\hat A^b_{in} \right]/\sqrt{2} $ and $\hat P_{in} = i\left[ (\hat A^b_{in}) ^\dagger - \hat A^b_{in} \right]/\sqrt{2}$ are the quadratures of the optical input of Bob's device, as defined in Eq.~\eqref{blueinout}. Here $\eta$ is the efficiency of the coherent displacement applied by Bob at the end of the protocol, and we defined the parameters $R =e^r - \eta \sqrt{e^{2r} -1}$ and $R^\prime =  \sqrt{e^{2r} -1} - \eta e^r$. This teleportation protocol is deterministic, the teleportation fidelity tends to 1 for very large squeezing $r\gg 1$ and  $\eta=1$. 

In Fig.~\ref{teleportF}(c) and (d), we show the fidelity of the final state  for different sets of state-of-the-art parameters~\cite{Krause15, Guha16}, in the special case in which a coherent state is teleported. Notice that in Eq.~\eqref{teleportstate} we neglected for the sake of simplicity the mechanical decoherence, yet a full treatment was used in Fig.~\ref{teleportF} and can be found in the Supplemental Material.
For uniformly distributed coherent states, it can be shown~\cite{Hammerer05} that the optimal classical strategy gives an average fidelity $\mathcal{F}=1/2$, providing a lower bound for the validation of the implementation of quantum teleportation. Another interesting bound is given by the no-cloning limit~\cite{Cerf00,Grosshans01} $\mathcal{F}>2/3$, which certifies that Bob has the best existing copy of the original state. Our analysis shows that these limits can be achieved with physical parameters available in state of the art experiments.

\begin{figure}[t]
\centering
\includegraphics[angle=0, width=0.45\textwidth]{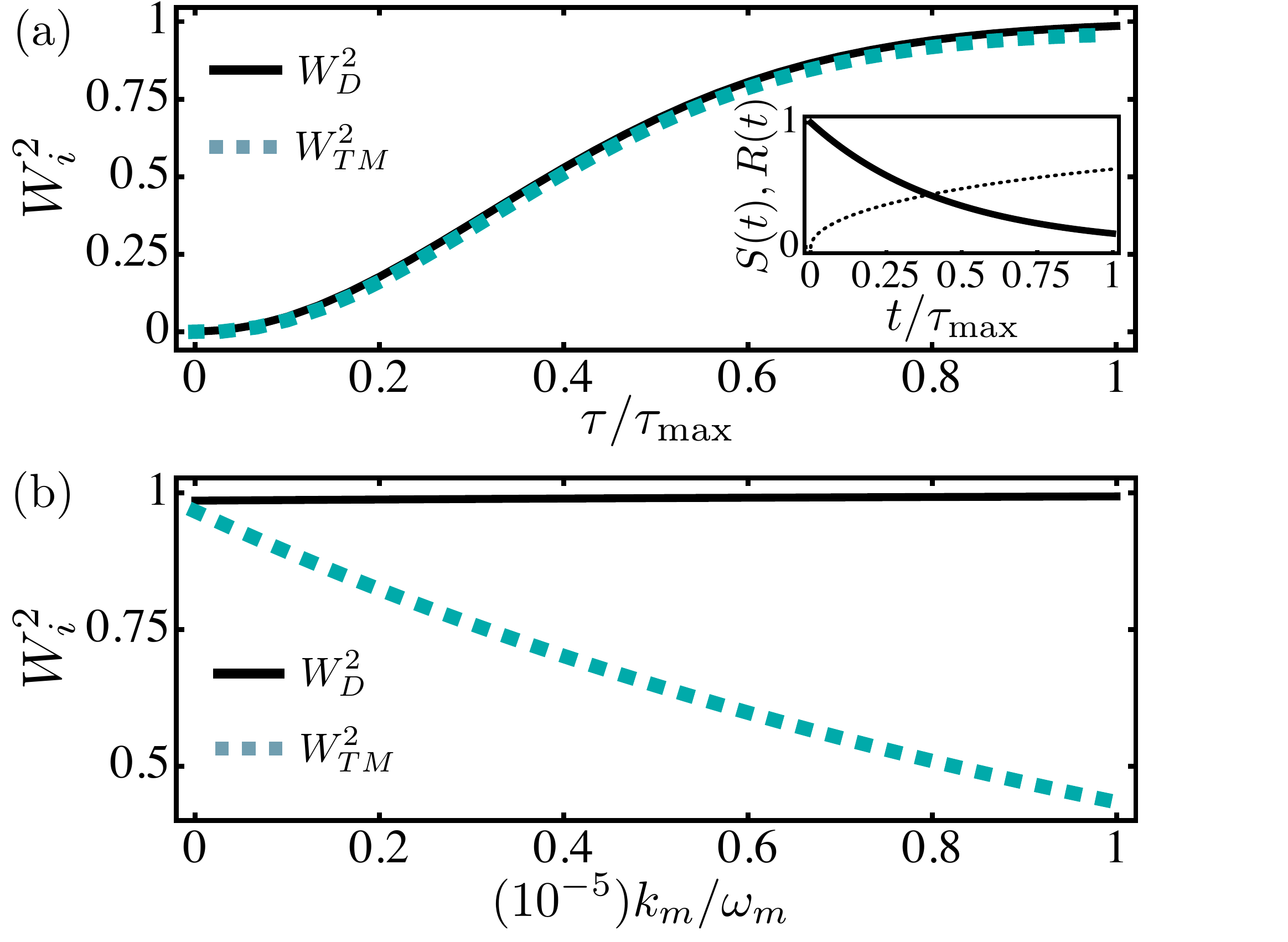}
\caption{\label{transferF} State transfer protocol. Squared weights of the initial state $\hat{m}(0)$ in several states of the protocol, as functions of the pulses duration $\tau$ (a), and of the mechanical dissipation rate $k_{m}$ (b).  The parameters are $\tau_{\text{max}} = 4 \times 10^3 / k_{c}$, $g=0.05 \, k_{c}$, $k_{c} = 0.1 \, \omega_{m}$, $\omega_{c} / k_{c}=10^{7}$, while $\omega_{m}/ 2\, \pi \sim 10^9$ Hz and $\omega_{c}/ 2\, \pi \sim 10^{15}$ Hz. (a) $k_{m}=\omega_{m} / 10^{7}$. (b) $\tau=\tau_{\text{max}}$. INSET : optimal temporal shape of the control pulses for the sender (full curve) and the receiver (dotted curve).}   
\end{figure}

Let us now consider the case in which a quantum state is directly exchanged between two remote mechanical resonators, as in Fig~\ref{protocols}.
With respect to quantum teleportation, quantum state transfer does not require entanglement sharing beforehand, nor feed-forward operations, at the cost of using a quantum communication channel and driving pulses of specific temporal shapes. The protocol is organised as follows: a red detuned driving pulse $\beta_{S} (t) = \beta S(t)$ of duration $\tau$ is sent through Alice's control port, mapping the state of her mechanical resonator onto a propagating light pulse $\hat a_{out} (t)$. The quantum signal is sent to Bob, who modulates the optomechanical interaction in his device via a  red-detuned control pulse $\beta_{R} (t) = \beta R(t)$. The interaction within Bob's toolbox completes the transfer between Alice's $\hat m_1 (0)$ and Bob's $\hat m_2 (\tau)$ mechanical resonators. Time-modulated classical drivings are needed to overcome a fundamental issue, absent in the teleportation protocol presented above. The photon emission and absorption processes are the time-reversal of each other, when flat drivings are considered. Hence, the emitted pulse sent by Alice would not have the right time-envelope to be efficiently absorbed by Bob. We have optimized, for arbitrary  states, the temporal shapes for the pumps used by the sender $S(t)$ and by the receiver $R(t)$, in order to maximize the transfer efficiency. The inset in Fig.~\ref{transferF}(a) shows the optimal functions $S(t)$ and $R(t)$ for the state-of-the-art parameters considered in our work (see Supplemental Material).

At the end of the protocol, the states of Alice's mechanical resonator and Bob's mechanical resonator are
\begin{eqnarray}
\hat m_1 (\tau) &=&  \sqrt{1-W_{D}^2} \, \hat m_1(0)+ N_1\big( \hat a_{in}, \hat m_{in}, \tau   \big), \label{mSfinal} \\
\hat m_2 (\tau) &=&   W_{TM} \,\hat m_1(0)  + N_2\big( \hat a_{in}, \hat m_{in}', \hat m_{in}, \tau   \big), \nonumber 
\end{eqnarray}
where for the sake of clarity we gathered all the contributions due to the optical and thermal environment in $N_1$ and $N_2$. The parameter $ W_{TM}$ establishes how significant is the transfer of $\hat m_1 (0)$ to $\hat m_2 (\tau)$, optimal state transfer corresponds to $ W_{TM}=1$. The parameter $W_{D}$ shows how the state $\hat m_1 (\tau)$ of the sender (Alice) is destroyed throughout the protocol.  In Fig.~\ref{transferF}, we show the square of these quantities  since it establishes the efficiency of the transfer for the second order moments. 
For state-of-the-art experimental parameters we approach the  limit $\hat m_2 (\tau) \approx \hat m_1(0)$,
which shows that the pulse sequence is optimal, i.e.,  there are no losses of information due to the time-envelope mismatch between the emission and absorption processes. Notice that, depending on the duration of the pump pulses, the state $\hat m_1 (0)$ can be transferred partially to Bob, and only partially destroyed in Alice's device. The proposed protocol allows to transfer an arbitrary quantum state in a tunable and non-destructive way.

In conclusion, we have introduced a novel interferometric scheme where the optomechanical coupling can be modulated in real time. This scheme enables us to design protocols to transfer arbitrary quantum states between remote mechanical resonators. The proposed design can be implemented with present technology, using optomechanical  crystals or whispering-gallery-mode
resonators~\cite{Ludwig12}. We considered state-of-the-art parameters of optomechanical systems working at telecom wavelength, but the proposed system could be straightforwardly adapted to superconducting microwave resonators~\cite{Lecocq15}.
Our work paves the way to the use of mechanical resonators as active nodes of quantum networks for continuous variable quantum information.
\begin{acknowledgments}
This work was supported by University Sorbonne Paris Cit\'e EQDOL and ANR COMB projects, and Brazilian Agencies CAPES (Proc 9498/13-3).
S.D. acknowledges the Institut Universitaire de France. I.F. acknowledges support of the ERC through the GANOMS project.
\end{acknowledgments}

\newpage

%%%_______________________________________supplemental material

\begin{widetext}

\section*{SUPPLEMENTAL MATERIAL}

\section{Derivation of the system Hamiltonian and coupling to the baths}
\label{Model_sup}

This section details the steps to go from the standard form of the system Hamiltonian (Eq.~(1) in the main text), to the form used throughout the paper (Eq.~(3) in the main text), by showing how the intracavity modes are coupled to the environment modes.

We start from the system Hamiltonian (Eq.~(1) in the main text),
\begin{equation}
\hat H =  \omega_c \left( \hat d^\dagger\hat d + \hat c^\dagger \hat c\right) + \omega_m \hat m^\dagger \hat m   - g_0 \left( \hat {d}^\dagger \hat d - \hat c^\dagger \hat c \right) \left( \hat m^\dagger + \hat m \right),
\label{H1_sup}
\end{equation}
where we denote with $\omega_c$ and $\omega_m$ the frequencies of the optical cavities and mechanical modes, respectively, and with $g_0$ the optomechanical coupling strength. To this Hamiltonian one has to add the coupling between the system and the environment,
\begin{align}
\hat{H}_{SE} & = i \, \int_{-\infty}^{\infty} d\omega \, \kappa_{c} (\omega) \Big( \hat{\gamma}_B^\dagger (\omega) \hat{c} (t) - \hat{\gamma}_B (\omega) \hat{c}^\dagger (t) \Big) + i \, \int_{-\infty}^{\infty} d\omega \, \kappa_{c} (\omega) \Big( \hat{\delta}_B^\dagger (\omega) \hat{d} (t) - \hat{\delta}_B (\omega) \hat{d}^\dagger (t) \Big) + \nonumber \\
& + i \, \int_{-\infty}^{\infty} d\omega \, \kappa_{m} (\omega) \Big( \hat{\mu}_B^\dagger (\omega) \hat{m} (t) - \hat{\mu}_B (\omega) \hat{m}^\dagger (t) \Big),
\label{HSE_sup}
\end{align}
and the free Hamiltonian of the baths
\begin{equation}
\hat{H}_{E} = \int_{-\infty}^{\infty} d\omega \, \omega \, \Big( \hat{\gamma}_B^\dagger (\omega) \hat{\gamma}_B (\omega) +  \hat{\delta}_B^\dagger (\omega) \hat{\delta}_B (\omega) + \hat{\mu}_B^\dagger (\omega) \hat{\mu}_B (\omega) \Big).
\label{Hbaths_sup}
\end{equation}
The full Hamiltonian of the system and the baths considered together is thus $\hat H_{TOT} = \hat H + \hat{H}_{SE} + \hat{H}_{E}$. Note that due to the Markov approximation, we neglect the frequency dependence of the system-baths coupling and use the notations $\kappa_{c} (\omega) = \sqrt{\kappa / \pi}$ and $\kappa_{m} (\omega) = \sqrt{\gamma / \pi }$.

The system we study is an interferometer, whose physical inputs, such as laser pulses or beams, correspond to the modes $\hat \alpha_B (\omega)$ and  $\hat \beta_B (\omega)$. They can be related to the modes inside the interferometer $\hat{\gamma}_B (\omega)$ and $\hat{\delta}_B (\omega)$ through the relations $\hat \alpha_B (\omega) = \left(\hat{\gamma}_B (\omega) - \hat{\delta}_B (\omega) \right)/\sqrt{2}$ and $\hat \beta_B (\omega) = \left(\hat{\gamma}_B (\omega) + \hat{\delta}_B (\omega) \right)/\sqrt{2}$. It is then meaningful to rewrite Eq.~(\ref{HSE_sup}) in terms of the operators $\hat \alpha_B (\omega)$ and  $\hat \beta_B (\omega)$, and we obtain
\begin{align}
\hat{H}_{SE} & = i \, \int_{-\infty}^{\infty} d\omega \, \kappa_{c} (\omega) \Big( \hat{\alpha}_B^\dagger (\omega) \frac{\left( \hat{c} (t) -  \hat{d} (t)  \right)}{\sqrt{2}} - \hat{\alpha}_B (\omega) \frac{\left( \hat{c}^\dagger (t) -  \hat{d}^\dagger (t)  \right)}{\sqrt{2}} \Big) + \nonumber \\
& + i \, \int_{-\infty}^{\infty} d\omega \, \kappa_{c} (\omega) \Big( \hat{\beta}_B^\dagger (\omega) \frac{\left( \hat{c} (t) +  \hat{d} (t)  \right)}{\sqrt{2}} - \hat{\beta}_B (\omega) \frac{\left( \hat{c}^\dagger (t) +  \hat{d}^\dagger (t)  \right)}{\sqrt{2}} \Big) + \nonumber \\
& + i \, \int_{-\infty}^{\infty} d\omega \, \kappa_{m} (\omega) \Big( \hat{\mu}_B^\dagger (\omega) \hat{m} (t) - \hat{\mu}_B (\omega) \hat{m}^\dagger (t) \Big),
\label{HSE2_sup}
\end{align}
where we can naturally define two collective intracavity modes $\hat a = \left(\hat c -\hat d \right)/\sqrt{2}$  and  $b = \left(\hat c +\hat d \right)/\sqrt{2}$, coupled to $\hat{\alpha}_B (\omega)$ and $\hat{\beta}_B (\omega)$ respectively.
By rewriting Eq.~(\ref{H1_sup}) in terms of the new defined collective optical modes, we obtain
\begin{equation}
\hat H =  \omega_c \left(\hat a^\dagger \hat a + \hat b^\dagger \hat b\right) + \omega_m \hat m^\dagger \hat m   + g_0 \left(\hat  a^\dagger \hat b +\hat a \, \hat b^\dagger \right)\left(\hat m^\dagger +\hat m \right). \label{H2_sup}
\end{equation}

We now consider that an undepleted coherent state at frequency $\omega_b$ is sent through the port $\hat \beta_{B}(\omega)$, which is essentially the input of the intracavity mode $\hat b$, such that this driven mode $\hat b$ is fairly approximated by a coherent state. We can then make the following replacement $\hat b\rightarrow \beta \, e^{-i \omega_{b} t}$, where $\beta$ is the size of the coherent state inside the cavity. The Hamiltonian becomes
\begin{equation}
\hat H \approx  \omega_c \hat a^\dagger \hat a + \omega_m \hat m^\dagger \hat m   + g_0 \left(\hat  a^\dagger \beta e^{-i \omega_{b} t} +\hat a \, \beta^\ast e^{i \omega_{b} t} \right)\left(\hat m^\dagger +\hat m \right). \label{H3_sup}
\end{equation}
To remove the time dependence from the Eq.~(\ref{H3_sup}) we use the unitary transformation $\hat U (t) = e^{i \omega_{b} \hat a^\dagger \hat a t}$ to move in a picture where the Hamiltonian is $\hat H_{\text{eff}} = \hat U \hat H \hat U^\dagger + i\frac{d\, \hat U}{dt} \hat U^\dagger$, which gives us the final expression of the effective Hamiltonian we use further on
\begin{equation}
\hat H_{\text{eff}} =  \left( \omega_c - \omega_b \right) \hat a^\dagger \hat a + \omega_m \hat m^\dagger \hat m   + g_0 \beta \left(\hat  a^\dagger +\hat a \right)\left(\hat m^\dagger +\hat m \right), \label{Hlin_sup}
\end{equation}
where we considered $\beta$ being real.

\section{Teleportation between two distant mechanical resonators}
\label{Teleportation_sup}

The teleportation protocol described here consists of three steps. First, an optomechanical entangled state is generated by Bob, who keeps one partie of the state (a mechanical resonator), and sends the other partie (an optical pulse) to Alice. The generation of this entangled EPR state is detailed in section \ref{EPR_sup}. Subsequently, Alice implements a state-swap process, $i.e.$, a beam splitter-type interaction between the light pulse coming from Bob and her own mechanical resonator, which is initially in the state to teleport. This part of the protocol is detailed in section \ref{StateSwap_sup}. Finally, Alice performs a Bell measurement of an optical quadrature and a mechanical quadrature, sends the outcomes through a classical channel to Bob, who implements a phase-space rotation and a phase-space displacement on the state of his mechanical resonator, which concludes the teleportation protocol. This last step is described in section \ref{MCC_sup}.

\subsection{EPR state generation}
\label{EPR_sup}

The first step of the teleportation protocol is the generation of an EPR state by Bob, between his own mechanical resonator, a double sided moving mirror, and a light pulse, which is sent to Alice after the entangling interaction. By working in a regime where the effective optomechanical coupling is much smaller than the cavity dissipation rate, the latter being itself much smaller than the mechanical frequency, namely $g_{0} \beta \ll \kappa  \ll \omega_{m}$, it is possible to enhance one of the two fundamentally different interactions appearing in the system Hamiltonian (\ref{Hlin_sup}). This first step requires to send a blue detuned pump pulse $\beta$ with a frequency $\omega_b = \omega_c + \omega_m$ to the bottom input port of Bob's interferometer, that enhances a two-mode squeezing interaction, thus reducing the system Hamiltonian, in a frame rotating with $\omega_{m}$, to  
 \begin{equation}
\hat H_I^b = g_0\beta
 \left( \hat a^\dagger \hat m^\dagger_2 + \hat a \, \hat m_2 \right),
\label{Hblue_sup}
\end{equation}
where $\hat m_2$ represents Bob's mechanical resonator, the one towards which the initial state of a distant mechanical resonator $\hat m_1$ will be teleported.

With the standard Input-Output theory \cite{Gardiner_sup,GardinerBook_sup,WallsBook_sup}, we take into account for both the optical and the mechanical dissipations, and derive the following Langevin equations
\begin{eqnarray}
\label{Entlan_sup}
\dot {\hat a}(t) &=& - \kappa   \hat a(t) -i g_{0} \beta \, \hat m_2^\dagger (t) - \sqrt{2 \kappa }\ \hat a_{in}(t), \\
\dot {\hat m}_2(t) &=&  - \gamma  \hat m_2(t) -i g_{0} \beta \, \hat a^\dagger (t)  - \sqrt{2 \gamma}\ \hat m_{in}(t). \label{Entlan2_sup}
\end{eqnarray}
The regime $g_{0} \beta \ll \kappa $ allows us to adiabatically eliminate the optical mode $\hat{a}$,
\begin{equation}
\label{Adiab_sup}
\hat a(t) \approx -i \frac{g_{0} \beta}{\kappa }\hat m_2^\dagger (t) - \sqrt{\frac{2}{\kappa }}\ \hat a_{in}(t),
\end{equation}
and to obtain expressions for $\hat{a}_{out} (t)$, the quantum optical output of the EPR generation toolbox, and $\hat{m}_2 (t)$, the time evolved annihilation operator of Bob's mechanical resonator
\begin{eqnarray}
\label{Fulla_sup}
\hat a_{out}(t) &=& - \hat a_{in}(t) - i \sqrt{2 \, G} e^{(G-\gamma )t} \hat m_2^\dagger(0) -  \sqrt{2  \, G} e^{(G-\gamma )t} \int_{0}^{t} dt' e^{-(G-\gamma )t'} \Big( \sqrt{2  \, G} \, \hat a_{in}^\dagger (t') - i\sqrt{2  \, \gamma } \, \hat m_{in}(t')\Big), \\
\hat m_2(t) &=&  e^{(G-\gamma )t}\hat m_2(0) +e^{(G-\gamma )t} \int_{0}^{t} dt' e^{-(G-\gamma )t'} \Big( i \sqrt{2  \, G} \, \hat a_{in}^\dagger (t') - \sqrt{2  \, \gamma } \, \hat m_{in}(t')\Big), \label{Fullm_sup}
\end{eqnarray}
with $\hat a_{out} (t) = \hat a_{in}(t) + \sqrt{2 \kappa }\hat a (t)$ and $G= (g_{0} \beta)^2 / \kappa $.

We will now make two important approximations in Eq. (\ref{Fulla_sup}) and (\ref{Fullm_sup}). We use the fact that in our model $\gamma  \ll G$, and first, take $G - \gamma  \approx G$, and second, neglect the effect of the mechanical bath on the optical mode, which gives us
\begin{eqnarray}
\label{Fulla2_sup}
\hat a_{out}(t) &\approx& - \hat a_{in}(t) - i \sqrt{2  \, G} \, e^{G \, t} \hat m^\dagger_2 (0) -  2  \, G \, e^{G \,t} \int_{0}^{t} dt' e^{-G \,t'} \hat a_{in}^\dagger (t'), \\
\hat m_2(t) &\approx&  e^{G \, t}\hat m_2 (0) +i \sqrt{2 \, G} \, e^{G \, t} \int_{0}^{t} dt' e^{-G \, t'} \hat a_{in}^\dagger (t') - \sqrt{2 \, \gamma } \, e^{G \, t} \int_{0}^{t} dt' e^{-G \, t'} \hat m_{in}(t'). \label{Fullm2_sup}
\end{eqnarray}
We now use the following definitions for normalized temporal field modes
\begin{eqnarray}
\label{Pdef_sup}
\mathit{P}(\tau,\hat O) = \sqrt{\frac{2G}{1-e^{-2G\tau}}}\int_0^\tau dt e^{-Gt}\hat O (t),\\
 \mathit{ Q}(\tau,\hat O ) = \sqrt{\frac{2G}{e^{2G\tau}-1}}\int_0^\tau dt e^{Gt}\hat O (t), \label{Qdef_sup}
\end{eqnarray}
to rewrite Eq. (\ref{Fulla2_sup}) and (\ref{Fullm2_sup}) into the convenient forms
\begin{eqnarray}
\label{blueainout_sup}
\hat A^b_{out} &=& -  e^{G \, \tau}\hat A^b_{in} - i\sqrt{ e^{2G  \, \tau}-1 }\ \left(\hat M_{in}\right)^\dagger, \\
\hat M_{out} &=& \ e^{G  \, \tau} \hat M_{in}  + i\sqrt{ e^{2G  \, \tau}-1 }\  \left(\hat A^b_{in}\right)^\dagger - C^b \hat M^b_B, \label{blueminout_sup}
\end{eqnarray}
where $\hat A^b_{in} =  \mathit{ P}(\tau,\hat a_{in} )$, $\hat A^b_{out} = \mathit{Q}(\tau,\hat a_{out} )$, $\hat M_{in} =  \hat m_2 (0)$, $\hat M_{out} =  \hat m_2 (\tau)$,  $\hat M^b_B =  \mathit{ P}(\tau,\hat m_{in} )$ and  $C^b = \sqrt{ ( e^{2G  \tau}-1 ) \gamma /G}$. 

Using the Eq. (\ref{blueainout_sup}) and (\ref{blueminout_sup}), the quadratures of these modes are
\begin{eqnarray}
\label{Xaout_sup}
\hat X_{out} &=& -  e^{r}\hat X_{in} - \sqrt{ e^{2r}-1 } \, \hat P_{2}(0), \\
\hat P_{out} &=& -  e^{r}\hat P_{in} - \sqrt{ e^{2r}-1 } \, \hat X_{2}(0), \label{Paout_sup} \\
\hat X_2 &=& e^{r}\hat X_2 (0) + \sqrt{ e^{2r}-1 } \, \hat P_{in} - C^b X^b_B, \label{X2_sup} \\
\hat P_2 &=& e^{r}\hat P_2 (0) + \sqrt{ e^{2r}-1 } \, \hat X_{in} - C^b P^b_B, \label{P2_sup}
\end{eqnarray}
where $r = G \, \tau$, $X_{in} =  \left[ (\hat A^b_{in}) ^\dagger +\hat A^b_{in} \right]/\sqrt{2} $ and $P_{in} = i\left[ (\hat A^b_{in}) ^\dagger - \hat A^b_{in} \right]/\sqrt{2}$ are the quadrature of the optical input of Bob's device, and where $X_{2} (0) =  \left[ \hat m_{2}^\dagger (0) +\hat m_{2} (0) \right]/\sqrt{2} $, $X_{2} =  \left[ \hat m_{2}^\dagger (\tau) +\hat m_{2} (\tau) \right]/\sqrt{2} $, $P_{2} (0) =  i \left[ \hat m_{2}^\dagger (0) - \hat m_{2} (0) \right]/\sqrt{2} $, and $P_{2} =  i \left[ \hat m_{2}^\dagger (\tau) - \hat m_{2} (\tau) \right]/\sqrt{2} $.

Eq. (\ref{Xaout_sup})-(\ref{P2_sup}) are similar to those of a standard optical two-mode squeezed state with a squeezing parameter $r$ \cite{Braunstein_sup}, except for the terms containing $X^b_B$, $P^b_B$, $i.e.$, the thermal noise of Bob's mechanical resonator. Their influence prohibit the standard property of a infinitly squeezed state in the case $r \rightarrow \infty$, since its increase conjointly grows $C^b$, which eventually kills the entanglement (see Fig. 2(a) in the main text). It is thus important to find the optimal point $r_{\text{opt}}$, for which the amount of entanglement is sufficient to carry out the teleportation protocol without a significant thermal disturbance.

By taking a thermal state for Bob's mechanical oscillator, $\langle \hat X_2 (0) \rangle = \langle \hat P_2 (0) \rangle = (n_{T} + 1/2)$, with $n_{T}$ being the thermal occupation number, and vacuum for Bob's optical quantum input, $\langle \hat X_{in} \rangle = \langle \hat P_{in} \rangle = 1/2$, the EPR entanglement criterion writes 
\begin{eqnarray}
\Delta_{\text{EPR}} &=& \left[ \Delta \Big( \hat X_{in} +\hat P_{2} \Big) \right]^2 + \left[ \Delta \Big( \hat P_{in} +\hat X_{2} \Big) \right]^2 \nonumber \\
&=& 2(n_{T} +1) \Big( e^{r} - \sqrt{ e^{2r}-1 } \Big)^2 + \frac{\gamma }{G} \Big(  e^{2r}-1 \Big) (2 \, n_{T} +1), \label{EPRcrit_sup}
\end{eqnarray}
where we took the mechanical initial state and the mechanical bath to be at thermal equilibrium. In Fig. 2(a) of the main text, the dashed curve is plotted using Eq. (\ref{EPRcrit_sup}), while the full curve is plotted using Eq. (\ref{EPRcrit_sup}) with $\gamma =0$, namely, a model where one neglects the mechanical bath contribution and thus retrieves a infinitly two-mode squeezed state for $r \rightarrow \infty$.

\subsection{State-swap, or beam splitter-type interaction}
\label{StateSwap_sup}

After Bob sent the pulse entangled with his mechanical resonator, Alice must implement a beam splitter-type interaction between this pulse and her mechanical resonator. This operation is carried out by pumping Alice's optomechanical toolbox from the bottom port with a red detuned drive pulse with a frequency $\omega_b = \omega_c - \omega_m$, considered here to have the same amplitude $\beta$, that enhances a beam splitter type interaction, thus reducing the system Hamiltonian, in a frame rotating with $\omega_{m}$, to  
 \begin{equation}
\hat H_I^r = g_0\beta
 \left( \hat a'^\dagger \hat m_1 + \hat a' \, \hat m_1^\dagger \right),
\label{Hred_sup}
\end{equation}
where $\hat m_1$ describes Alice's mechanical resonator, the state of which is to be teleported to Bob.

Using the same method as before, we derive the Langevin equations for Alice's toolbox
\begin{eqnarray}
\label{Redlan_sup}
\dot {\hat a}'(t) &=& - \kappa   \hat a'(t) -i g_{0} \beta \, \hat m_1 (t) - \sqrt{2 \kappa }\ \hat a_{in}'(t), \\
\dot {\hat m}_1(t) &=&  - \gamma  \hat m_1(t) -i g_{0} \beta \, \hat a' (t)  - \sqrt{2 \gamma}\ \hat m_{in}'(t). \label{Redlan2_sup}
\end{eqnarray}
We again adiabatically eliminate the optical mode and obtain the evolution for $\hat{a}_{out}' (t)$ and $\hat{m}_1 (t)$
\begin{eqnarray}
\label{FullaRed_sup}
\hat a_{out}'(t) &=& - \hat a_{in}'(t) - i \sqrt{2 \, G} e^{-(G+\gamma )t} \hat m_1 (0) +  \sqrt{2  \, G} e^{-(G+\gamma )t} \int_{0}^{t} dt' e^{(G+\gamma )t'} \Big( \sqrt{2  \, G} \, \hat a_{in}' (t') + i \sqrt{2  \, \gamma } \, \hat m_{in}'(t')\Big), \\
\hat m_1 (t) &=&  e^{-(G+\gamma )t}\hat m_1(0) +e^{-(G+\gamma )t} \int_{0}^{t} dt' e^{(G+\gamma )t'} \Big( i \sqrt{2  \, G} \, \hat a_{in}' (t') - \sqrt{2  \, \gamma } \, \hat m_{in}'(t')\Big), \label{FullmRed_sup}
\end{eqnarray}
We use again the fact that $\gamma  \ll G$, take $G - \gamma  \approx G$, and neglect the effect of the mechanical bath on the optical mode,
\begin{eqnarray}
\label{Fulla2Red_sup}
\hat a_{out}'(t) &\approx& - \hat a_{in}'(t) - i \sqrt{2  \, G} \, e^{-G \, t} \hat m_1 (0) +  2  \, G \, e^{-G \,t} \int_{0}^{t} dt' e^{G \,t'} \hat a_{in}' (t'), \\
\hat m_1(t) &\approx&  e^{-G \, t}\hat m_1 (0) +i \sqrt{2 \, G} \, e^{-G \, t} \int_{0}^{t} dt' e^{G \, t'} \hat a_{in}' (t') - \sqrt{2 \, \gamma } \, e^{-G \, t} \int_{0}^{t} dt' e^{G \, t'} \hat m_{in}'(t'). \label{Fullm2Red_sup}
\end{eqnarray}
We write Eq. (\ref{Fulla2Red_sup}) and (\ref{Fullm2Red_sup}) into the convenient forms
\begin{eqnarray}
\label{redainout_sup}
\hat A^r_{out} &=& -  e^{-Gt}\hat A^r_{in} - i\sqrt{ 1-  e^{-2Gt} }\ \hat M_{in}', \\
\hat M_{out}' &=& \ e^{-Gt} \hat M_{in}'  + i\sqrt{ 1-  e^{-2Gt}  }\  \hat A^r_{in} - C^r \hat M^r_B, \label{redminout_sup}
\end{eqnarray}
where $\hat A^r_{in} =  \mathit{ Q}(\tau,\hat a_{in}' )$, $\hat A^r_{out} = \mathit{P}(\tau,\hat a_{out}' )$, $\hat M_{in}' =  \hat m_1 (0)$, $\hat M_{out}' =  \hat m_1 (\tau)$,  $\hat M^r_B =  \mathit{ Q}(\tau,\hat m_{in} )$ and  $C^r = \sqrt{ ( 1 - e^{- 2G  \tau} ) \gamma /G}$. Note that here $\hat a_{in}' = \hat a_{out}$, meaning that the quantum optical output of Bob's toolbox is the quantum optical input of Alice's toolbox. Hence, we have $\hat A^r_{in} = \hat A^b_{out}$, meaning that Bob's output pulse shape perfectly matches Alice's optical input shape.
Using the Eq. (\ref{redainout_sup}) and (\ref{redminout_sup}), the quadratures of these modes are
\begin{eqnarray}
\label{Xu_sup}
\hat X_u &=& -  e^{-r'}\hat X_{out} + \sqrt{ 1 - e^{-2r'} } \, \hat P_{1}(0), \\
\hat P_u &=& -  e^{-r'}\hat P_{out} - \sqrt{ 1 - e^{-2r'} } \, \hat X_{1}(0), \label{Pu_sup} \\
\hat X_v &=& - e^{-r'}\hat X_1 (0) - \sqrt{ 1 - e^{-2r'} } \, \hat P_{out} - C^r X^r_B, \label{Xv_sup} \\
\hat P_v &=& - e^{-r'}\hat P_2 (0) + \sqrt{ e^{-2r}-1 } \, \hat X_{out} - C^r P^r_B, \label{Pv_sup}
\end{eqnarray}
where $r'= G \, \tau'$, $X_{out} =  \left[ (\hat A^b_{out}) ^\dagger +\hat A^b_{out} \right]/\sqrt{2} $ and $P_{out} = i\left[ (\hat A^b_{out}) ^\dagger - \hat A^b_{out} \right]/\sqrt{2}$, where $X_{u} =  \left[ (\hat A^r_{out}) ^\dagger +\hat A^r_{out} \right]/\sqrt{2} $ and $P_{u} = i\left[ (\hat A^r_{out}) ^\dagger - \hat A^r_{out} \right]/\sqrt{2}$, and where $X_{1} (0) =  \left[ \hat m_{1}^\dagger (0) +\hat m_{1} (0) \right]/\sqrt{2} $, $X_{v} =  \left[ \hat m_{1}^\dagger (\tau') +\hat m_{1} (\tau') \right]/\sqrt{2} $, $P_{1} (0) =  i \left[ \hat m_{1}^\dagger (0) - \hat m_{1} (0) \right]/\sqrt{2} $, and $P_{v} =  i \left[ \hat m_{1}^\dagger (\tau') - \hat m_{1} (\tau') \right]/\sqrt{2} $. 

One can see that in Eq. (\ref{Xu_sup})-(\ref{Pv_sup}), a judicious choice of $\tau'$ can lead to a 50-50 beam splitter-type interaction. Indeed, by choosing $\tau'$ such that $e^{-r'} = 1/\sqrt{2}$, Eq. (\ref{Xu_sup})-(\ref{Pv_sup}) become
\begin{eqnarray}
\label{Xu2_sup}
\hat X_u &=& \frac{ \hat P_{1}(0) -\hat X_{out} }{\sqrt{2}}, \\
\hat P_u &=& \frac{ -  \hat X_{1}(0) -\hat P_{out}}{\sqrt{2}}, \label{Pu2_sup} \\
\hat X_v &=& \frac{- \hat X_1 (0) - \hat P_{out} }{\sqrt{2}} - \sqrt{\frac{\gamma}{2G}} X^r_B, \label{Xv2_sup} \\
\hat P_v &=& \frac{ - \hat P_2 (0) +  \hat X_{out}}{\sqrt{2}} - \sqrt{\frac{\gamma}{2G}} P^r_B, \label{Pv2_sup}
\end{eqnarray}

We now use Eq. (\ref{X2_sup}) and (\ref{Xv2_sup}) to write Bob's mechanical resonator $\hat{X}_{2}$ quadrature as
\begin{eqnarray}
\hat X_2 &=&  \big( \hat X_2 + \hat P_{out} \big) + \sqrt{2} \hat X_v + \hat X_1 (0) + \sqrt{\frac{\gamma}{G}} X^r_B, \nonumber \\
&=& \hat X_1 (0) + \Big( e^{r} - \sqrt{ e^{2r}-1 } \Big) \hat X_2 (0) + \Big( \sqrt{ e^{2r}-1 } - e^{r} \Big) \hat P_{in} + \sqrt{2} \hat X_v +  \sqrt{\frac{2 \, \gamma}{G}} X^r_B - C^b X^b_B. \label{X2rep_sup}
\end{eqnarray}
By using Eq. (\ref{P2_sup}) and (\ref{Xu2_sup}) we can write Bob's mechanical resonator $\hat{P}_{2}$ quadrature as
\begin{eqnarray}
\hat P_2 &=&  \big( \hat P_2 + \hat X_{out} \big) + \sqrt{2} \hat X_u - \hat P_1 (0), \nonumber \\
&=& - \hat P_1 (0) + \Big( e^{r} - \sqrt{ e^{2r}-1 } \Big) \hat P_2 (0) + \Big( \sqrt{ e^{2r}-1 } - e^{r} \Big) \hat X_{in} + \sqrt{2} \hat X_u  - C^b X^b_B. \label{P2rep_sup}
\end{eqnarray}

\subsection{Measurement and classical channel}
\label{MCC_sup}

The next step of the protocol is the measurement of both $\hat X_v$ and $\hat X_u$ by Alice, which become classical random variables $X_v$ and $X_u$ after the measurement. Bob's mechanical resonator state collapses into a state which differs from the intial Alice's mechanical resonator state by a phase-space rotation and a phase-space displacement, in a simplified case where one neglect the mechanical bath effects and takes $r \rightarrow \infty$.

Through the classical channel, Bob receives Alice's outcomes $X_v$ and $X_u$ and performs both the needed phase-space rotation and displacement,
\begin{eqnarray}
\label{X2tel_sup}
\hat X_2 \longrightarrow \hat X_2^{tel} &=& \hat X_2 - \eta \sqrt{2} \hat X_v, \\
\hat P_2 \longrightarrow \hat P_2^{tel} &=& -\hat P_2 + \eta \sqrt{2} \hat X_u, \label{P2tel_sup}
\end{eqnarray}
where the parameter $\eta$ describes both the efficiency of the measurement and of the displacement. 
We can rewrite the teleported state quadratures from Eq. (\ref{X2tel_sup}) and (\ref{P2tel_sup}) as
\begin{eqnarray}
\label{X2tel2_sup}
\hat X_2^{tel} &=& \eta \hat X_1 (0) + \Big( e^{r} - \eta \sqrt{ e^{2r}-1 } \Big) \hat X_2 (0) + \Big( \sqrt{ e^{2r}-1 } - \eta e^{r} \Big) \hat P_{in}  +  \eta \sqrt{\frac{2 \, \gamma}{G}} X^r_B - C^b X^b_B, \\
\hat P_2^{tel} &=& \eta \hat P_1 (0) - \Big( e^{r} - \eta \sqrt{ e^{2r}-1 } \Big) \hat P_2 (0) - \Big( \sqrt{ e^{2r}-1 } - \eta e^{r} \Big) \hat X_{in} + C^b X^b_B. \label{P2tel2_sup}
\end{eqnarray}
One can see that without mechanical dissipation, $i.e.$, $\gamma =0$, and with the limits $r \rightarrow \infty$ and $\eta \rightarrow 1$, one obtains the perfect teleportation with $X_2^{tel} = X_1 (0)$ and $P_2^{tel} = P_1 (0)$.

In order to check how successful the protocol is, we use the teleportation fidelity $\mathcal{F} = \langle \psi_{1} \vert \hat \rho_{2}^{tel}  \vert \psi_{1}  \rangle$, where $ \vert \psi_{1}  \rangle$ is the state to teleport, $i.e.$, the initial state of Alice's double sided moving mirror, and where $\rho_{2}^{tel}$ is the density matrix of the teleported state, $i.e.$, the final state of Bob's double sided moving mirror. We consider the initial state of Alice's mechanical resonator to be a coherent state with a displacement $\alpha_{1} = \left( X_{1} (0) + i \, P_{1} (0)\right)/\sqrt{2}$. In that case, $\mathcal{F} = \pi Q_{tel} (\alpha_{1})$, where $Q_{tel}$ is the $Q$ function of the teleported state \cite{Braunstein_sup},
\begin{equation}
\mathcal{F} = \frac{1}{2 \sqrt{\sigma_{X} \sigma_{P}}} \exp{\Bigg[ -\left( 1 - \eta \right)^2 \Bigg( \frac{X_{1}^2 (0)}{2 \sigma_{X}}  + \frac{P_{1}^2 (0)}{2 \sigma_{P}}\Bigg) \Bigg]}.
\label{Ftel_sup}
\end{equation}
$\sigma_{X}$ and $\sigma_{P}$ are the variances of the $Q$ function,
\begin{eqnarray}
\sigma_{P} &=& \sigma_{X} + \eta^2 \frac{\gamma  }{G} \bigg( n_{T} + \frac{1}{2} \bigg) \nonumber \\
&=&  \frac{1}{4} \big( 1 + \eta^2\big) +\frac{1}{4}\Big( \sqrt{ e^{2r}-1 } - \eta e^{r} \Big)^2  +\Bigg[ \frac{1}{2}  \Big( e^{r} - \eta \sqrt{ e^{2r}-1 } \Big)^2 + \frac{(C^b)^2}{2}  + \eta^2 \frac{\gamma  }{G} \Bigg] \bigg( n_{T} + \frac{1}{2} \bigg),
\label{sigmas_sup}
\end{eqnarray}
where we used the fact that initially, Bob's double sided moving mirror, and the mechanical baths of both mechanical resonators are all at thermal equilibrium.

Eq.  (\ref{X2tel_sup}) and (\ref{P2tel_sup}) imply that the measurement of both the mechanical and the light quadratures $\hat X_v$ and $\hat X_u$ have been performed with the same precision. Let us us briefly comment this statement. For these two measurements, Alice proceeds as follows: the optical quadrature $\hat X_u$ is measured with a homodyne scheme applied to Alice's output pulse, which concludes this first measurement. Subsequently, an additional red detuned pulse $\beta$ with a different duration $\tau$ is sent through the pumping port of Alice's toolbox, in order to transfer the mechanical quadrature $\hat X_v$ to a second output light pulse, which quadratures will be measured again with a homodyne measurement. To clarify this point, let us write the expressions of Alice's optical output quadratures, after the second pulse of duration $\tau''$
\begin{eqnarray}
\label{XMEAS_sup}
\hat X_{meas} &=& -  e^{-r''}\hat X_{OVN} + \sqrt{ 1 - e^{-2r''} } \, \hat P_{v}, \\
\hat P_{meas} &=& -  e^{-r''}\hat P_{OVN} - \sqrt{ 1 - e^{-2r''} } \, \hat X_{v}, \label{PMEAS_sup}
\end{eqnarray}
where $\hat X_{meas}$ and $\hat P_{meas}$ are the quadratures of the second quantum optical output, and where $\hat X_{OVN}$ and $\hat P_{OVN}$ are the quadratures of the quantum optical vacuum noise (OVN) involved as an input in the interaction process cause by the second red detuned pulse.

By choosing $\tau''$ such that $e^{-r''} \rightarrow 0$, Eq. (\ref{XMEAS_sup}) and (\ref{PMEAS_sup}) become
\begin{eqnarray}
\label{XuMEAS_sup}
\hat X_{meas} &=& \hat P_v, \\
\hat P_{meas} &=&  - \hat X_v, \label{PuMEAS_sup}
\end{eqnarray}
thus showing that both $\hat X_v$ and $\hat X_u$ can be measured with the same precision.

\section{Tunable quantum state transfer between two distant mechanical resonators}
\label{TQST_sup}

The principle of the tunable quantum state transfer protocol is to transfer the state $\hat m_1$ of a mechanical resonator inside the same interferometric toolbox as used before, to $\hat m_2$, the state of a second mechanical resonator inside a second interferometer, spatially distant from the first one. Essentially, the final goal of the protocol is very similar to the one of the teleportation protocol discussed above, yet using different means. Here, no shared entanglement is required, nor the collapse of a state due to measurements, as means to bring $\hat m_2$ into the target state for the teleportation. Additionally, no classical communication is needed, however a quantum channel is essential for the protocol detailed here.

In fact, the procedure itself of the tunable quantum state transfer is similar to the teleportation protocol, taking place as follows: the first toolbox, containing the state to transfer, is pumped by a classical driving pulse at the bottom vertical port. This yields in the generation of a quantum output pulse coming out of the horizontal port and flying towards the second toolbox. While this flying pulse enters the second toolbox by its horizontal port, the latter is pumped by a classical driving pulse at its bottom vertical port. The interaction taking place in the second toolbox leads to the tunable transfer of the first mechanical state towards the second mechanical resonator. 

There are two major differences in these two steps, with respect to the teleportation protocol. The first difference is that both toolboxes are pumped with a red detuned classical driving pulses. The second major difference is that the use of the same temporal pulse shapes as before does not lead to a significant transfer efficiency. Indeed, in the teleportation protocol the fact that Bob's toolbox was pumped by a blue detuned drive made its quantum output pulse shape perfect to be absorbed by Alice's toolbox, which was pumped at the same time by a red detuned drive. The fact that here both toolboxes are pumped by red detuned drives makes the quantum output pulse of the first toolbox very difficult to absorb by the second toolbox. 

To overcome this complication one needs to optimise the temporal shapes of both red detuned pumps. We shall now call the toolboxes as the sender and the receiver, which pump pulses, both of duration $\tau$, have the following amplitudes $\beta_{S} (t) = \beta S(t)$ and $\beta_{R} (t) = \beta R(t)$, where
\begin{eqnarray}
\label{St_sup}
S (t) &=& \sqrt{1-e^{-\mu_{S} G t}}, \\
R (t) &=& e^{-\mu_{R} G t} \label{Rt_sup}
\end{eqnarray}
are functions specifying the pulse shapes (see inset in Fig. 3(a) in the main text), and where $G= (g_{0} \beta)^2 / \kappa $. $\mu_{S}$ and $\mu_{R}$ are parameters that optimise the transfer efficiency. For $g_{0} \beta =0.05 \, \kappa $, we have $\mu_{S}=0.05$ and $\mu_{R}=0.22$.

Both pumps enhance a state-swap process, described by the Hamiltonian (\ref{Hred_sup}), which gives the following Langevin equations for the sender, similar to Eq. (\ref{Redlan_sup}) and (\ref{Redlan2_sup}),
\begin{eqnarray}
\label{LanSa_sup}
\dot {\hat a}(t) &=& - \kappa   \hat a(t) -i g_{0} \beta S(t) \, \hat m_1 (t) - \sqrt{2 \kappa }\ \hat a_{in}(t), \\
\dot {\hat m}_1(t) &=&  - \gamma  \hat m_1(t) -i g_{0} \beta S(t) \, \hat a (t)  - \sqrt{2 \gamma}\ \hat m_{in}(t). \label{LanSm_sup}
\end{eqnarray}
As in the previous section, we adiabatically eliminate the optical mode and obtain the evolution for $\hat{a}_{out} (t)$ and $\hat{m}_1 (t)$
\begin{eqnarray}
\label{aOUTS_sup}
\hat a_{out}(t) &=& - \hat a_{in}(t) - i \sqrt{2 \, G} S(t) e^{-\gamma t - G \int_{0}^{t} dt' S^2(t')} \hat m_1 (0) + \nonumber \\
&+&  \sqrt{2  \, G} S(t) e^{-\gamma t - G \int_{0}^{t} dt' S^2(t')} \int_{0}^{t} dt' e^{\gamma t' + G \int_{0}^{t'} dt'' S^2(t'')} \Big( \sqrt{2  \, G} S(t') \, \hat a_{in} (t') + i \sqrt{2  \, \gamma } \, \hat m_{in}(t')\Big), \\
\hat m_1 (t) &=&  e^{-\gamma t - G \int_{0}^{t} dt' S^2(t')} \hat m_1(0) + \nonumber \\
&+& e^{-\gamma t - G \int_{0}^{t} dt' S^2(t')} \int_{0}^{t} dt' e^{\gamma t' + G \int_{0}^{t'} dt'' S^2(t'')} \Big( i \sqrt{2  \, G} S(t') \, \hat a_{in} (t') - \sqrt{2  \, \gamma } \, \hat m_{in}(t')\Big). \label{mS_sup}
\end{eqnarray}
We now need the Langevin equations for the process occurring for the receiver, not very different from Eq. (\ref{LanSa_sup}) and (\ref{LanSm_sup}),
\begin{eqnarray}
\label{LanRa_sup}
\dot {\hat a}'(t) &=& - \kappa   \hat a'(t) -i g_{0} \beta R(t) \, \hat m_2 (t) - \sqrt{2 \kappa }\ \hat a_{in}'(t), \\
\dot {\hat m}_2(t) &=&  - \gamma  \hat m_2(t) -i g_{0} \beta R(t) \, \hat a' (t)  - \sqrt{2 \gamma}\ \hat m_{in}'(t). \label{LanRm_sup}
\end{eqnarray}
Accordingly, the expressions for $\hat{a}_{out}' (t)$ and $\hat{m}_2 (t)$ are similar to the Eq. (\ref{aOUTS_sup}) and (\ref{mS_sup}),
\begin{eqnarray}
\label{aOUTR_sup}
\hat a_{out}'(t) &=& - \hat a_{in}'(t) - i \sqrt{2 \, G} R(t) e^{-\gamma t - G \int_{0}^{t} dt' R^2(t')} \hat m_2 (0) + \nonumber \\
&+&  \sqrt{2  \, G} R
(t) e^{-\gamma t - G \int_{0}^{t} dt' R^2(t')} \int_{0}^{t} dt' e^{\gamma t' + G \int_{0}^{t'} dt'' R^2(t'')} \Big( \sqrt{2  \, G} R(t') \, \hat a_{in}' (t') + i \sqrt{2  \, \gamma } \, \hat m_{in}'(t')\Big), \\
\hat m_2 (t) &=&  e^{-\gamma t - G \int_{0}^{t} dt' R^2(t')} \hat m_2(0) + \nonumber \\
&+& e^{-\gamma t - G \int_{0}^{t} dt' R^2(t')} \int_{0}^{t} dt' e^{\gamma t' + G \int_{0}^{t'} dt'' R^2(t'')} \Big( i \sqrt{2  \, G} R(t') \, \hat a_{in}' (t') - \sqrt{2  \, \gamma } \, \hat m_{in}'(t')\Big), \label{mR_sup}
\end{eqnarray}
where the quantum input pulse of the sender is the quantum output pulse of the receiver, $i.e.$, $\hat a_{in}' (t) = \hat a_{out} (t)$. Using Eq. (\ref{mR_sup}) and (\ref{aOUTS_sup}) we obtain
\begin{eqnarray}
\hat m_2 (t) &=&  e^{-\gamma t - G \int_{0}^{t} dt' R^2(t')} \hat m_2(0) + 2 \, G e^{-\gamma t - G \int_{0}^{t} dt' R^2(t')}    \int_{0}^{t} dt' R(t') S(t') e^{ G \int_{0}^{t'} dt'' \left( R^2(t'') - S^2(t'') \right) } \hat m_1(0) + \nonumber \\
&+& N\big( \hat a_{in}, \hat m_{in}', \hat m_{in}, t   \big), \label{mRfinal_sup}
\end{eqnarray}
Where $N\big( \hat a_{in}, \hat m_{in}', \hat m_{in}, t   \big)$ encodes the dependence on the vacuum noise of the quantum optical input of the sender, and the thermal noises of both the sender and the receiver.

We can rewrite the final states of Eq. (\ref{aOUTS_sup}), (\ref{mS_sup}), and (\ref{mRfinal_sup}), $i.e.$, after the end of the interaction time $\tau$, into the useful forms
\begin{eqnarray}
\label{aOUTSfinal_sup}
\hat a_{out}(\tau) &=& W_{TA} \, \hat m_1 (0) + \dots, \\
\hat m_1 (\tau) &=&  \sqrt{1-W_{D}^2} \, \hat m_1(0)+ \dots, \label{mSfinal_sup} \\
\hat m_2 (\tau) &=&   W_{TM} \,\hat m_1(0)  + \dots, \label{mRfinal2_sup}
\end{eqnarray}
where we omitted for clarity all the other contributions different than those of the initial state intended for the transfer, $\hat m_1 (0)$. The three quantities defined in Eq. (\ref{aOUTSfinal_sup}), (\ref{mSfinal_sup}), and (\ref{mRfinal2_sup}) give us the following information: $W_{TA}$ shows if it is possible to obtain information on $\hat m_1 (0)$ simply by measuring the pulse travelling between the sender and the receiver ; $W_{D}$ shows how the state $\hat m_1 (\tau)$ of the sender is destroyed by the transfer of its initial state $\hat m_1 (0)$ to the receiver's state $\hat m_2 (\tau)$ ; $ W_{TM}$ shows how significant is the transfer of $\hat m_1 (0)$ to $\hat m_2 (\tau)$, thus defining a benchmark for the protocol. We show the square of these quantities in Fig. 4 in the main text since it demonstrates the efficiency of the transfer for the second order moments of the initial state.

\end{widetext}
\end{document}